# A Hybrid Trim Strategy for Coaxial Compound Helicopter

Yuan Su*, Zeyuan Wang**, Yihua Cao***

*Beihang University, 100191, Beijing, People's Republic of China*

| Article info | Abstract |
|---|---|
| Keywords:<br>Coaxial compound helicopter<br>Trim strategy<br>Optimization<br>Elevator control<br>Control redundancy | Interest in the coaxial compound helicopter (CCH) has been increasing in the civil aviation and engineering community for its high-speed and high-maneuverability features, and is likely to continue to do so for the foreseeable future. Since the control in CCH is totally different from the conventional helicopter, the redundant control strategy design is one of the biggest challenges. In this study, the CCH model based on XH-59A is built to investigate the impact of the propeller and the elevator on the flight performance. Four trim strategies with different objectives are proposed and then compared to find the optimal control allocation. A heuristic descent search method is applied to search the optimal velocity at which the propeller and the elevator are engaged. A significant improvement of power required at medium and high speed with acceptable rotor airloads increment was found by using the Hybrid Trim strategy in the speed range of 0-100m/s, with regard to a pre-configured pitch angle schedule. The corresponding control variables obtained locate in a reasonable control range, with a maximum power reduced of 13% at 100m/s, which showcases the potential of the Hybrid Trim strategy. |

## 1. Introduction

The coaxial compound high-speed helicopter [1] (abbreviated as CCH) is a type of helicopter with two coaxial rotors and auxiliary propulsion. The most significant advantage of a CCH is its high-speed ability and excellent maneuverability. The maximum speed of a conventional helicopter is about 70m/s (252km/s, 140knots), while the CCH XH-59A developed by Sikorsky and NASA in the 1980s reached 135m/s (487km/h, 270knots) [2]. In 2005, the high-speed helicopter research project and the X2 Demonstrator were launched by Sikorsky, performing the first flight in August 2008, and created a speed record of 128m/s (463km/h, 256knots) in a flight test in September 2010 [3][4].

Compared with a conventional helicopter, the CCH could overcome a series of problems such as retreating blade stall, advancing blade compressibility, rotor vibration, etc., but it also brings challenges in flight control. The CCH replaces the tail rotor with two coaxial counter-rotating rotors to provide the lift and to balance the rotor torques. In addition, a pusher propeller is added to the tail of the CCH, providing extra forward thrust at high speed. The CCH is also equipped with movable elevators and rudders, constituting different blade control strategies from conventional helicopters: main rotor collective pitch, differential collective, longitudinal and lateral cyclic pitch, longitudinal and lateral differential, propeller collective, elevator, and rudder control. The prerequisite for studying the flight control problem of CCH is to establish a stable trim flight state. However, if the trim variables were found only through the six-degree-of-freedom equations, there are infinite solutions, so-called the redundant problem. In this case, some restrictions must be considered, such as preset variables or optimal conditions.

The trim problem of the CCH has been carried out by many scholars in related work. A pitch angle preset schedule was proposed by K. Ferguson and D. Thomson in the literature [5], which compares the aerodynamic characteristics of CCH and HCH (hybrid compound helicopter). The two authors also analyzed the performance characteristics of CCH in detail in the document [6] and found that the pusher propeller engaged above 50m/s (100 knots) would significantly affect the performance. Researchers Y H. Zhang, X. Zhao built the model based on XH-59A in the literature [7] to investigate the control phase angle limit and compared the simulation results with the real flight test data. Y. Yuan, D. Thomson in the literature [8] studied the effect of different propeller control strategies on trim and flight quality and found that the propeller control strategy would affect the flight performance and range. These two authors also study the impact of the elevator on the handling quality and maneuverability in the literature [22]. For high-speed helicopters of other configurations (perhaps the most famous model is the UH-60 equipped with propulsive propeller and wings), many scholars have also studied their trim strategies, including the "Flight to Optimal" control strategy [9], surrogate model-based optimization using response surface model [10], and even using the artificial intelligence (neural network model) [11]. Almost all researches take the required power into account as an essential indicator of trim strategy. It is worth noting that there are still few literatures that individually investigated the elevator on trim, especially for allocating the propeller and the elevator control to achieve a new optimal state.

This article focuses on the CCH trim problem and study in detail the control allocation of the propeller and the elevator operating simultaneously. The results have proved that the elevator control strategy is vital for CCH's required power, the distribution of forces and moments, and the rotor airloads. The results are of great significance for designing the control system.

In the Methodology section, the CCH modeling process will be introduced in detail, including the trim process and the preliminary analysis of the elevator. Four configurations of elevator control (BL, STrim, MPTrim, HTrim) and the corresponding optimization algorithms are then proposed based on the preliminary analysis. In the Results section, the trim results produced by these four strategies are analyzed and compared with their advantages and disadvantages. A new concept is proposed called Tolerance of Increased Load, as a constraint and an indicator of the side effect brought by the elevator. The conclusion part sums up the article. The most important results obtained are the order and the optimal velocities to engage the propeller and the elevator, and the fact that HTrim, the recommended optimal strategy proposed, can be used to reduce the total required power with an acceptable rotor airloads increase.

## 2. Methodology

### 2.1 Coaxial compound helicopter modeling

The CCH in this paper is a compound helicopter with two rigid coaxial main rotors and a pusher propeller on the tail. Its parameters are shown in **Table 1**. The definition of each control variable, and the positive directions as well as their feasible range are shown in

**Table 2**. Based on these parameters, this section will elaborate on the mathematical modeling of each component.

2.1.1 Coaxial rotors modeling

The CCH in this study uses the same rotors as the ones of XH-59A, which has two hingeless counter-rotating coaxial rotors. The hingeless rotor can be equivalent to a hinged rotor [12], represented by **Fig. 1**.

The relation between the flapping frequency, the dimensionless equivalent hinge offset, the stiffness, the flap moment of inertia, and the first moment of inertia of a hingeless rotor and a hinged rotor, is shown in equation (1). The dimensionless equivalent hinge offset calculated based on the parameters in **Table 1** is 0.47.

---

\* Associate Professor, School of Aeronautic Science and Engineering, Beihang University, No. 37 Xueyuan Road, Haidian District, Beijing 100191, People's Republic of China.

\*\* Corresponding author: Master candidate, École Centrale de Pékin, Beihang University, No. 37 Xueyuan Road, Haidian District, Beijing 100191, People's Republic of China.

*** Professor, School of Aeronautic Science and Engineering, Beihang University, No. 37 Xueyuan Road, Haidian District, Beijing 100191, People's Republic of China.

*E-mail addresses:* suyuan@buaa.edu.cn (Y. Su) zeyuan_wang@outlook.com (Z Y. Wang) yihuacaobu@126.com (Y. Cao),

# Nomenclature

| | | | | |
|---|---|---|---|---|
| $A$ | Rotor disk area……………………………………….. m² | | $\delta_r$ | Rudder deflection angle…………………………....... deg |
| $e$ | Non-dimensional equivalent flapping hinge offset | | $\delta_{l2u}, \delta_{u2l}$ | Rotor inflow interference factor |
| **F,M** | Vectors of forces and moments…………………….. N, N.m | | $\varphi$ | Roll angle……………………………………………. deg |
| $K_\beta$ | Stiffness of the blade…………………………………. N.m/rad | | $\psi$ | Azimuth position/yaw angle..……………………… deg |
| $K_{1s}, K_{1c}$ | First harmonic coefficients in inflow model | | $\theta$ | Pitch angle…………………………………………… deg |
| $I_\beta$ | flap moment of inertia……………………………….. kg.m² | | $\theta_0$ | Main rotor collective pitch ……………..………….. deg |
| $I_y$ | Moment of inertia in y-direction of body axes………… kg.m² | | $\theta_{1c}$ | Main rotor lateral cyclic pitch……………..………... deg |
| $M_y$ | Pitch moment……………………………………….. N.m | | $\theta_{1cdiff}$ | Main rotor lateral differential……………..……….. deg |
| $M_\beta$ | First moment of inertia………………………………….. kg.m | | $\theta_{1s}$ | Main rotor longitudinal cyclic pitch……………….…. deg |
| $r_b$ | Radial position…………………………………………….. m | | $\theta_{1sdiff}$ | Main rotor longitudinal differential………………..…. deg |
| $R, R_{prop}$ | Radius of main rotor and propeller…………………… m | | $\theta_{diff}$ | Main rotor differential collective…………………….. deg |
| $T_u, T_l$ | Upper and lower rotor thrust……………………….. N | | $\theta_{prop}$ | Propeller collective pitch angle.…………………….. deg |
| $u_u, v_u, w_u,$ $u_l, v_l, w_l$ | Translational velocities of upper and lower rotor hub in body axes…………………………………………….. m/sec | | $\omega_n$ | Flapping frequency……………….….…………….. rad/sec |
| $v_{i,u}, v_{i,l}$ | Upper and lower rotor induced velocities……………. m/sec | | $\Omega$ | Main rotor rotational speed…………..…………….. rad/sec |
| $v'_{i,u}, v'_{i,l}$ | Upper and lower rotor inherent induced velocities…….. m/sec | | $\chi$ | Main rotor wake skew angle……………….………… deg |
| $v'_i$ | Inherent induced velocity…………………………….. m/sec | | $\mu$ | Advance ratio |
| $v'_{i0}$ | Non-periodic term in inherent induced velocity……….. m/sec | | $\lambda$ | Non-dimensional rotor inflow |
| **x** | Vector of trim variables | | $\rho$ | Air density………………………………………….... kg/m³ |
| $\delta_e$ | Elevator deflection angle……………………………… deg | | | |

**Table 1**

Definition of general parameters

| Parametric variables | Values |
|---|---|
| Mass (kg) | 5500 |
| Rotor radius (m) | 5.49 |
| Number of blades | 2x3 |
| Rotational speed (rad/s) | 35 |
| Spring stiffness (N.m/rad) | 220500 |
| Rotor solidity | 0.127 |
| Shaft tilt (deg) | 3 |
| Rotor blade twist gradient (deg) | -10 |
| Flap moment of inertia (kg.m²) | 450 |
| Flapping frequency (Ω) | 1.4 |
| Lock number | 5.41 |
| Lower rotor hub position (m) | (0.00, 0.00, -0.89) |
| Shaft spacing (m) | 0.77 |
| Number of propeller blades | 4 |
| Propeller blade airfoil | Clark Y |
| Propeller radius (m) | 1.3 |
| Propeller rotational speed (rad/s) | 162 |
| Propeller blade twist gradient (deg) | -30 |
| Propeller solidity | 0.2 |
| Propeller hub position (m) | (-7.66 0.00, 0.00) |
| Horizontal stabilizer position | (-6.80, 0.00, 0.20) |
| Vertical stabilizer position | (-6.8, 0.00, -0.50) |

**Table 2**

Definition of control variables

| Control variables | Positive Direction/Definition | Range (deg) |
|---|---|---|
| $\theta_0$ | Blade pitch angle increases | [0,20] |
| $\theta_{diff}$ | $\theta_{diff} = \dfrac{\theta_{0,u} - \theta_{0,l}}{2}$ | [-5,5] |
| $\theta_{1c}$ | Swashplat tilted down in the left | [-6.25,6.25] |
| $\theta_{1cdiff}$ | $\theta_{1cdiff} = \dfrac{\theta_{1c,u} - \theta_{1c,l}}{2}$ | [0,4.5] |
| $\theta_{1s}$ | Swashplat tilted down in the back | [-10,10] |
| $\theta_{1sdiff}$ | $\theta_{1sdiff} = \dfrac{\theta_{1s,u} - \theta_{1s,l}}{2}$ | [-1,1] |
| $\theta_{prop}$ | Blade pitch angle increases | [0,70] |
| $\delta_e$ | Trailing edge down | [-25,25] |
| $\delta_r$ | Trailing edge left | [-30,30] |

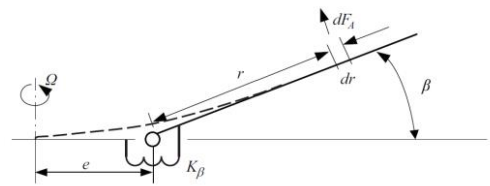

**Fig. 1** Schematic diagram of rigid rotor flapping equivalence

$$\omega_n = \Omega \sqrt{1 + \frac{eM_\beta}{I_\beta} + \frac{K_\beta}{I_\beta \Omega^2}} \quad (1)$$

In order to establish the mathematical model of the main rotors, we use Pitt-Peters dynamic inflow model, blade element theory, and momentum theory. For the upper and lower rotors, the induced velocity is the superposition of its own inherent induced velocity and the interference from another rotor, which is written in equation (2):

$$\begin{cases} v_{i,u}(\psi) = v'_{i,u}(\psi) + \delta_{l2u} v'_{i,l}(\psi) \\ v_{i,l}(\psi) = v'_{i,l}(\psi) + \delta_{u2l} v'_{i,u}(\psi) \end{cases} \quad (2)$$

Where $v_{i,u}$ (resp. $v_{i,l}$) is the induced velocity of the upper (resp. lower) rotor, $\psi$ is the azimuth position of the blade, $\delta_{l2u}$ (resp. $\delta_{u2l}$) is the inflow interference factor generated from the lower (resp. upper) rotor to the upper (resp. lower) rotor, and $v'_{i,u}$ (resp. $v'_{i,l}$) is the inherent induced velocity of the upper (resp. lower) rotor calculated by the Pitt-Peters inflow model. The Pitt-Peters inflow model in the form of first-order harmonics is represented in equation (3) in hub wind axes.

$$\begin{cases} v'_i(\psi, r_b) = v'_{i0} + \frac{r_b}{R} v'_{i0}(K_{1s} \sin(\psi) + K_{1c} \cos(\psi)) \\ K_{1s} = 0 \\ K_{1c} = \frac{15\pi}{32} \tan\left(\frac{\chi}{2}\right) \end{cases} \quad (3)$$

Where $\chi$ is the rotor wake skew angle, $K_{1s}$ and $K_{1c}$ are the first-order harmonic coefficients of the induced velocity, which are the empirical functions of the rotor wake skew angle, representing the inhomogeneity of the inflow.

The determination of the interference factor is complicated and challengeable. In general, the interference factor measures the influence of one rotor on another. If no interference was considered, the $\delta_{u2l}$ and $\delta_{l2u}$ would both be 0. Obviously, the upper and lower rotors have different impact on each other. Since the lower rotor is inside the downwash area of the upper rotor, the interference produced by the upper rotor is stronger than the lower rotor, that is to say, $\delta_{u2l} \geq \delta_{l2u}$. According to the literature [13], the study takes $\delta_{u2l} = 1$ and $\delta_{l2u} = 0$ for the reason that the rotors are so close that the wake of the upper rotor has not yet fully developed before reaching the lower rotor. Therefore, the impact of the lower rotor is neglected. Indeed, the effect of the rotor model based on this assumption behaves still relatively well in the literature [5][8][13]. However, through the CFD simulation and verification, the document [14] found that $\delta_{l2u}$ and $\delta_{u2l}$ are not constantly 0 and 1, but are functions of the advance ratio, as shown in **Fig. 2**. The biggest interference between the rotors appears in the hover state, but it is greatly reduced and tends towards 0 in the high-speed forward flight. This is mainly due to the fact that the wake of the rotor is almost parallel to the rotor disk at high speed (the rotor wake skew angle is almost 90°), and few parts of the wake of the upper rotor spreads to the lower rotor, so the interference is significantly reduced, or even converges to 0 if the velocity tends to infinite. This model is used in this article, considering that it is more credible for the real flight, with the validation in the literature [15][16], rather than taking a simple constant of 0 or 1.

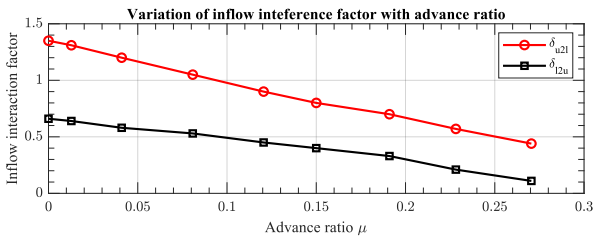

**Fig. 2** Variation of inflow interference factor with advance ratio

According to the inflow model, the induced velocity of each rotor at each flight speed is demonstrated in **Fig. 3**. It can be found that the result well represents the characteristics of the rotors. In general, because the lower rotor bears stronger interference from the upper rotor, its induced velocity is relatively larger. When flying at low speed, the induced velocity constitutes the main part of the rotor inflow to produce thrust. On the contrary, at high speed, the interference between rotors tends to be zero, and the inflow mainly consists of the forward airflow, which leads to the induced velocity converging to zero.

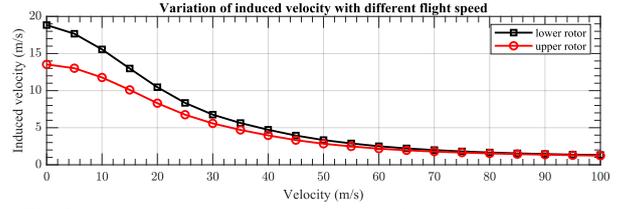

**Fig. 3** Variation of induced velocity with different flight speed

Blade element theory is a straightforward approach to compute the forces and moments on the rotor that is widely used in the literature [5][6][7][8][13][14][17]. The idea is common and direct: divide each blade into finite elements, and then calculate and sum up (integration) the resultant force and moment on the small element. The integral force and moment are decomposed into harmonic components using the Fourier decomposition method. Since higher-order components mainly result in the vibration and acoustic issues, the terms affecting flight dynamics are primarily composed of first-order harmonics. Moreover, the flapping equation and its first-order harmonics are also obtained from the moment equation of the blade[17].

One of the necessary information used in blade element method is the induced velocity, and it would remain unknown if we could not have extra equations, i.e., the momentum equations. They provide a bridge between the blade element method and the inflow model through the relationship in equation (4).

$$\begin{cases} T_u = 2\rho A v'_{i0,u} \sqrt{u_u^2 + v_u^2 + (v_{i0,u} - w_u)^2} \\ T_l = 2\rho A v'_{i0,l} \sqrt{u_l^2 + v_l^2 + (v_{i0,l} - w_l)^2} \end{cases} \quad (4)$$

$T_u$ (resp. $T_l$) is the total thrust of the upper (lower) rotor. $v'_{i0,u}$ (resp. $v'_{i0,l}$) is the non-periodic term of the inherent induced velocity of the upper (resp. lower) rotor (using Fourier transform). $v_{i0,u}$ (resp. $v_{i0,l}$) is the non-periodic term of the total induced velocity of the upper (lower) rotor (using Fourier transform). The induced velocity could be iteratively calculated with other trim variables.

2.1.2 Propeller modeling

The tail pusher propeller for auxiliary thrust is the most important feature of the CCH, by which the forward thrust is generated to resist fuselage drag during high-speed flight. Some assumptions could facilitate the modelling that the propeller is considered as a fixed speed rigid rotor without flapping or cyclic pitch. The thrust is controlled by the collective pitch of the propeller blade. The forces and moments are calculated using the same method as the main rotors [13]. The corresponding parameters are given in **Table 1**.

2.1.3 Fuselage modelling

The fuselage model comes from the wind tunnel data of the helicopter Lynx [18], a model well-validated in the literature [5][6][8][13] to represent the fuselage of the CCH with excellent results. The fuselage constitutes a significant part of drag, especially when flying at high speed, and affects the dynamic performance of the helicopter.

2.1.4 Horizontal stabilizer, vertical stabilizer, elevator and rudder modelling

The modeling of the horizontal stabilizer, the vertical stabilizer, the elevator, and the rudder come from the wind tunnel data of XH-59A in the document [19], where the aerodynamic coefficients were measured with rotor turned on in trimmed state, so the aerodynamic interference effect of these components due to rotor wake is included in the data. The aerodynamic results manifest that the elevator and the rudder have little aerodynamic effect below the flight speed 40m/s. This could be explained as the wake of main rotors at low speed leading to the tail in the stall area. As the speed increases above 45m/s, the effect of rotor wake gradually decreases, so the control efficiency of stabilizers increases. The elevator has an important impact on trim, which is the main focus of this article and is analyzed in detail in the following sections.

*2.2 Trimming strategy*

The trim range of the CCH is limited below 100m/s. The helicopter in the trim state has no acceleration in any direction, formulated in equation (5).

$$\begin{cases} \sum \mathbf{F}(\mathbf{x}) = \mathbf{F}_U(\mathbf{x}) + \mathbf{F}_L(\mathbf{x}) + \mathbf{F}_{Prop}(\mathbf{x}) + \mathbf{F}_{Fus}(\mathbf{x}) + \mathbf{F}_{HS}(\mathbf{x}) + \mathbf{F}_{VS}(\mathbf{x}) + \mathbf{F}_G(\mathbf{x}) = \mathbf{0} \\ \sum \mathbf{M}(\mathbf{x}) = \mathbf{M}_U(\mathbf{x}) + \mathbf{M}_L(\mathbf{x}) + \mathbf{M}_{Prop}(\mathbf{x}) + \mathbf{M}_{Fus}(\mathbf{x}) + \mathbf{M}_{HS}(\mathbf{x}) + \mathbf{M}_{VS}(\mathbf{x}) + \mathbf{M}_G(\mathbf{x}) = \mathbf{0} \end{cases} \quad (5)$$

Forces and moments are combined of different parts of upper rotor (U), lower rotor (L), pusher propeller (Prop), fuselage (Fus), horizontal stabilizer (HS), vertical stabilizer (VS), and gravity (G). **X** is the trimmed variable defined as $\mathbf{x} = \left[\theta_0, \theta_{diff}, \theta_{1c}, \theta_{1s}, \theta, \varphi, \theta_{prop}, \delta_e, \delta_r, \theta_{1cdiff}, \theta_{1sdiff}\right]^T$.

The vectors of forces and moments expressed in the equation (5) include the gravity and all the aerodynamic forces and moments of the main rotors, pusher propeller, fuselage, vertical and horizontal stabilizer. These two vector equations could be expanded to six scalar equations in X, Y, and Z directions of body axes, corresponding to a set of solutions of six unknowns.

Note that **X** consists of 11 unknowns, so equation (5) is impossible to be directly solved to obtain one and only one solution. Therefore, additional equations are needed such as preset values of some variables or introducing optimization objectives. We set the longitudinal differential to 0, and the lateral differential is configured to limit the lift offset (LOS) value [13][20] so that the main rotors are in an efficient non-stall working condition. LOS is defined by the lateral lift-offset equation:

$$LOS = \frac{\Delta M_x}{TR} \quad (6)$$

Where $\Delta M_x$ is the differential rotor roll moment and $T$ is the total rotor lift. According to the literature[4], the LOS value can be set quadratically with the airspeed, which is $LOS = 0.0002V^2$. This LOS strategy is designed to improve the flight dynamics characteristics and performance across the flight range and is also employed in the literature [5][6][8][13][22][23][24].

For the trim problem, the rudder deflection angle could be set to 0. The aerodynamic efficiency of the rudder begins to be predominant at medium and high speeds to control the yaw moment and replace the main rotor differential. However, because of the little interference between the main rotors, the corresponding thrust and torque tend to be the same and the differential is near 0, so considering the rudder control strategy to replace the differential control is unnecessary for the static trim problem. Nevertheless, the rudder might be taken into account if the maneuverability is the objective, since the control efficiency of the rudder at medium and high speed is substantial. Contrary to the rudder, the elevator has a crucial impact on trim, especially at high speeds, producing longitudinal forces and considerable pitch-moments by which the required power and rotor airloads are highly influenced. The optimal elevator control will be discussed in detail in the following sections.

The research on the thrust setting of the tail propeller has been carried out by many scholars. One of the approaches is to set the pitch angle to a fixed value. Nevertheless, this is not always desirable because of the unnecessary power consumed at low speed [5]. Another approach is to configure an adaptive pitch angle [5], that is, the value of the pitch attitude at each speed is computed in consideration of a power optimization schedule, producing a closer attitude to level-flight at high-speed to improve the flight experience of pilots. The third method is a multi-objective optimization with regard to the optimal power and the flight quality [8]. It is worth noting that the elevator control is not considered in these three strategies mentioned above. In this article, we decide to use the second strategy with a pre-configured pitch angle schedule shown in **Fig. 4**, rather than the third strategy since the elevator is the main concern and is needed to be preliminarily studied to investigate more deeply the flight quality.

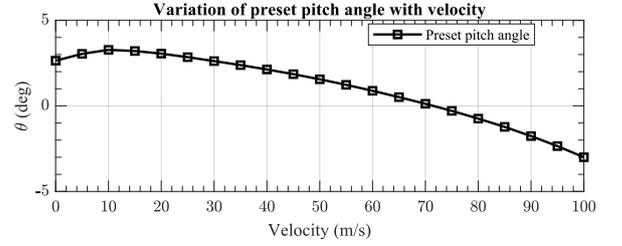

**Fig. 4** Variation of preset pitch angle with velocity

In summary, the trimmed variable **X** can be written as

$$\mathbf{x} = \left[\theta_0, \theta_{diff}, \theta_{1c}, \theta_{1s}, \varphi, \theta_{prop}, \theta = preset, \delta_e = preset, \delta_r = 0, \theta_{1sdiff} = 0, \theta_{1cdiff} = f(LOS)\right]^T$$

Retracting the unknown terms and we obtain $\mathbf{x} = \left[\theta_0, \theta_{diff}, \theta_{1c}, \theta_{1s}, \varphi, \theta_{prop}\right]^T$.

As a result, the 11 unknown variables can be reduced to 6, so that equation (5) has a unique solution.

The approach to solve the system of nonlinear equation (5) with constraints of each variable has been studied in the past years. Combined with modern computer science, we can conveniently implement the algorithms into programming software such as MATLAB and Python to quickly solve the problem. This article uses the Levenberg-Marquardt algorithm (L-M) [21], a method to convert the system of equations into a constraint optimization problem and then uses gradient descent to find the local optima.

L-M is not a global search method and highly depends on the choice of the initial point. Attempts might be made several times to change the initial value. Moreover, the closer the initial value to the solution, the higher the success probability of the algorithm. This paper solves all the solutions from the hover state to forward flight at 100m/s with an interval of 1m/s. The adopted strategy is to first try different initial values to obtain the solution in the hover state and then iterate the program in such a way: we set the initial value at the corresponding speed as the solution of the previous speed. Since these two values are very close, it can greatly improve the success probability and the speed of the search. For example, the initial value for solving 1m/s is the solution at 0m/s, and the initial value for solving 2m/s is the solution at 1m/s. The iteration procedure continues until 100m/s.

*2.3 The impact of elevator on required power and rotor load*

Flight speed is a key factor to determine the control efficiency of the elevator which can be represented as $CE_{elevator} = \Delta M_y / (\Delta \delta_e I_y)$ [20], where $\Delta M_y$ is the variation of the pitch moment with a small $\Delta \delta_e$. **Fig. 5** depicts the variation of $CE_{elevator}$ with velocity. At low speed, the control efficiency of the elevator is basically zero. Above 40-45m/s, the magnitude of $CE_{elevator}$ increases significantly. Therefore, one could conclude that when the speed is below 40m/s, the elevator has no effect on trim. And if the speed is greater than 40-45m/s, the trim state begins to be increasingly affected by the elevator. Especially at high speeds, the elevator will produce a considerable pitch moment.

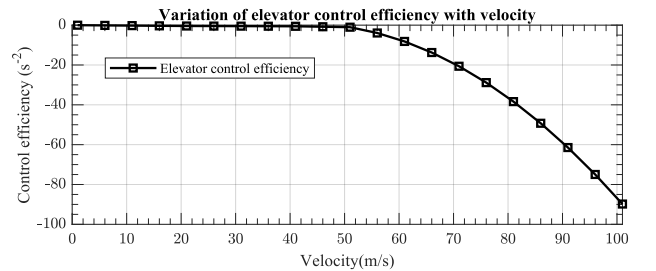

**Fig. 5** Variation of elevator control efficiency with velocity

**Fig. 6** shows the influence of the elevator on the trimmed required power and the main rotor airloads at several typical speeds (at 0m/s, 10m/s, 30m/s, 45m/s, 60m/s, 80m/s and 100m/s), where each point corresponds to a trimmed state with a specific airspeed and elevator deflection angle. The data at same speed are represented by the same shape and color. Among them, the solid patterns serve as the benchmark state where the deflection angle of the elevator is 0. The dashed lines across multiple data series are the isolines of elevator deflection angle.

It can be seen from the figure that at or below 40m/s, the elevator has no effect on required power and main rotor airloads (i.e., a single data point). But starting from 45m/s, since the control efficiency of the elevator gradually increases, the trimmed data diverges to several different points. As the elevator deflection angle decreases (trailing edge down), the required power decreases, and the main rotor airloads increases. After reaching a certain speed (the turning points), the required power increases slightly. **Table 3** presents a set of values in the high-speed mode (100m/s). Compared with the benchmark, an elevator deflection of -10° could reduce the required power by 54.2% with a 26.1% increase in the rotor airloads.

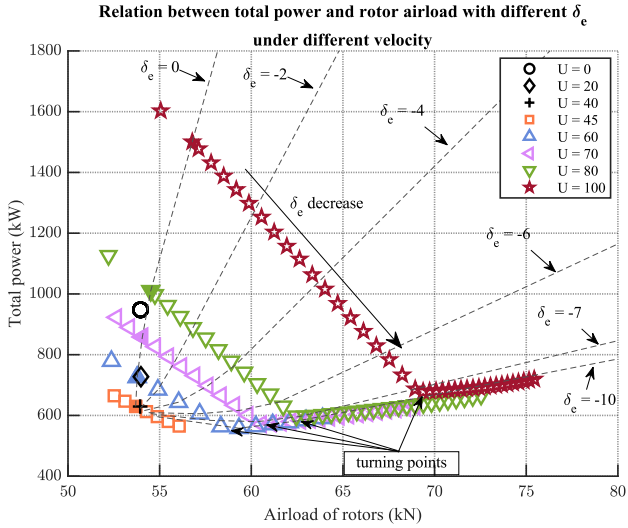

**Fig. 6** Variation of power required and rotor airloads with different elevator deflection

It should be noted that the turning point corresponds to the minimum power, after which the power starts to increase again, while the main rotor airloads increases monotonously. Indeed, the reduction of required power is favorable, while the increase of rotor airloads is undesirable. Because the rotor airloads affects the vibration, noise, structure and other characteristics of the CCH, and excessive load reduces the life cycle of the rotor, and proposes the new requirements for the design of the airfoil and material of the blade and the rotor hub. In consequence, a favorable elevator control strategy should take the required power as the performance indices and also the constraint of rotor load. In the literature[25], the author carefully examines the effects of lift offset on rotor blade airloads and structural loads, and point out that understanding and quantify the blade loading capability of coaxial lift-offset rotor is an important step to ultimately establish the blade design guidance for maneuver capability. Engineer should know the max airloads of the blade to well choose the material, airfoil and design the shape of the blade.

**Table 3**

Variation of required power and rotor load with different elevator angle under 100m/s

| Elevator deflection angle (deg) | Power required (reduced percentage) | Rotor load (increased percentage) |
| --- | --- | --- |
| 0 | 1501kW (baseline) | 56.8kN (baseline) |
| -2 | 1278kW (14.9%) | 60.2kN (6.0%) |
| -4 | 1039kW (30.8%) | 63.7kN (12.1%) |
| -6 | 804kW (46.4%) | 67.1kN (18.1%) |
| -7 | 687.1kW (54.2%) | 68.9kN (21.3%) |
| -10 | 687.2kW (54.2%) | 71.6kN (26.1%) |

*2.4 The impact of elevator on the distribution of forces and moments*

In order to explain the mechanism of the influence brought by the elevator on the required power and the rotor airloads, we analyze the impact of the elevator on the distribution of forces and moments of different components, represented in **Fig. 7** for a typical velocity of 70 m/s. In general, a more negative elevator deflection angle produces a greater pitch-up moment. Therefore, it is necessary to control the longitudinal cyclic pitch so that the rotor provides the corresponding pitch-down moment for equilibrium. At the same time, the thrust and the power of the propeller are reduced. This process also leads to the extra forces in x-direction to compensate for the drag generated by the fuselage, so the total thrust of the rotor increases.

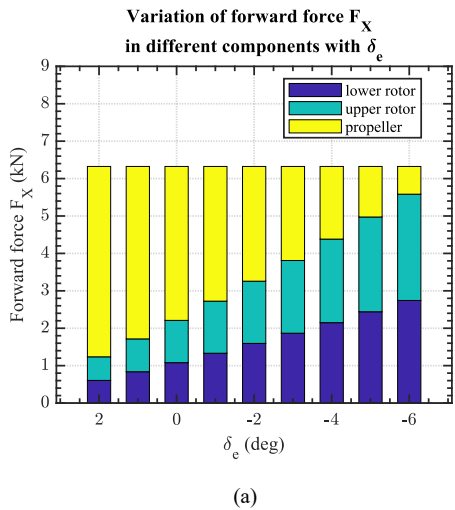

(a)

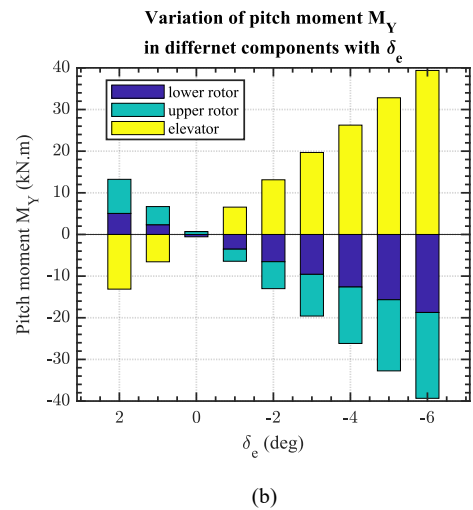

(b)

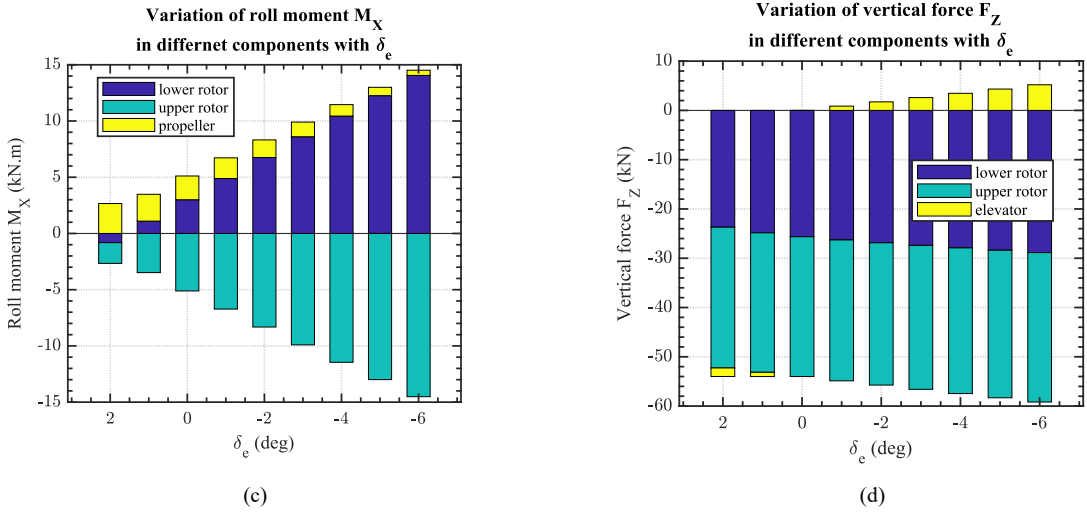

**Fig. 7** Variation of distribution of force and moment with different elevator deviation angle at 70 m/s

Looking further, at the flying speed of 70m/s, since the pitch angle is preset, the forces on the vertical tail and the fuselage (including gravity) are only related to the attitude and are not affected by the elevator, so the components in figures only include the main rotors, the propeller, and the elevator. Therefore, if one tries to calculate the total force or moments, the residual value is not 0 because it should be balanced with other force or moment of the vertical stabilizer and the fuselage, which are not presented in those figures.

It can be seen from **Fig. 7**(a) that as the deflection angle of the elevator decreases, the propeller is gradually unloaded while the forward force of the upper and lower rotors increases. **Fig. 7**(b) reflects the distribution of the pitch moments in each component. When the elevator deflection angle is positive, a pitch-down moment is generated, which is balanced with the main rotor's pitch-up moment. Conversely, as the elevator deflection angle is negative, a pitch-up moment is produced, while the main rotors generate pitch-down moments. Notice that, in **Fig. 7**(b) at $\delta_e = 0$, the pitch moments of the upper and the lower rotor are 0.69kN.m and -0.56kN.m respectively so it is not clearly illustrated. **Fig. 7**(a) and **Fig. 7**(b) demonstrate the co-effect of the pitch-down moment and forward force of main rotors resulting in the unloading of the propeller and the increase of the main rotor airloads. This can be used to explain the influence of the elevator on the rotor airloads in **Fig. 6**.

**Fig. 7**(c) shows the indirect effect of the elevator on the roll moment distribution, which reflects the fact that as the deflection angle of the elevator decreases, the roll moments of the main rotors increase and are opposite to each other. The lateral differential should be considered if the LOS generated by the roll moments is too large. Since a more negative deflection of the elevator leads to a lower forward force of propeller in **Fig. 7**(a), it also results in a decreased torque (roll moment) in **Fig. 7**(c) which is correlated positively with the pusher force. In consequence, the distribution of roll moments in CCH's components could be highly affected by the elevator, even though it does not directly produce the roll moment.

It is worth noting that although the elevator also generates lift, **Fig. 7**(d) demonstrates that this vertical force is tiny and could be negligible compared to other forces and moments.

*2.5 Pre-configured strategy of elevator*

In the trim strategy section, we only discussed the preset schedule of pitch angle but not the elevator deflection angle. In this section, different elevator control strategies are proposed based on the analysis in the previous section, including an optimal strategy adapted to the minimum power and rotor airloads, which we refer to as a Hybrid Trim strategy.

2.5.1 Strategy 0: baseline (BL)

The baseline trim (BL) disables both the propeller and the elevator control. This configuration is equivalent to a conventional coaxial helicopter, with a speed range limited in 0-70m/s. It serves as a benchmark to be compared with other strategies. In order to be fully compared in high-speed state, we extend the mathematical solution to 100m/s which is represented by the black dashed line in the figures. Readers need to keep in mind that the solution between 70m/s and 100m/s under this configuration has no practical significance, but only for the uniformity of plots and facilitation of comparison.

2.5.2 Strategy 1: Simple Trim (STrim)

The STrim strategy disables and set the elevator to 0 ($\delta_e = 0$). The advantage of this configuration is the simplicity of the control. The pitch attitude of the CCH will be fully determined by the longitudinal cyclic pitch and the propeller thrust. However, it also limits the potential of the elevator. The mathematical description of the trim problem is shown in (7)

$$\text{Given } U \in [0,100], \text{ Find } \mathbf{x} \text{ such that } \begin{cases} \sum \mathbf{F(x)} = \mathbf{0} \\ \sum \mathbf{M(x)} = \mathbf{0} \end{cases} \text{ subject to } \begin{cases} \theta(U) = \theta_{preset}(U) \\ \delta_e = 0 \end{cases} \quad (7)$$

2.5.3 Strategy 2: Minimum Power Trim (MPTrim)

Minimum Power Trim (MPTrim) is designed to engage the elevator if any reduction of required power can be achieved without constraint, formulated in (8).

$$\text{Given } U \in [0,100], \text{ Find } \delta_e = \arg\min_{\delta_e, \mathbf{x}} Power(\delta_e, \mathbf{x}) \text{ subject to } \begin{cases} \sum \mathbf{F(x)} = \mathbf{0} \\ \sum \mathbf{M(x)} = \mathbf{0} \\ \theta(U) = \theta_{preset}(U) \end{cases} \quad (8)$$

The advantage of MPTrim is evident that the required power can be maximumly reduced. However, some potential dangers must be considered. According to **Fig. 6**, the elevator will increase the main rotor airloads while reducing the power. For instance, when flying at the highest speed of 100m/s, MPTrim leads to the thrust of the main rotor increasing from 56.8kN to 71.6kN (see **Table 3**). In this case, the issues of vibration, retreating blade stall, and advancing blade compressibility would arise, posing a huge challenge to the flight safety of the CCH.

2.5.4 Strategy 3: Hybrid trim strategy (HTrim)

The last strategy Hybrid Trim (HTrim) is proposed as an optimal candidate which is a compromise between STrim and MPTrim. The elevator could be engaged if there is any possibility to reduce the power, with the constraint of limited rotor airloads. We proposed a new function *TIL* : Tolerance of Increased Load, defined as

$$TIL(U, \delta_e) = \left( \frac{T_{rotor}(U, \delta_e)}{T_{rotor}(U, \delta_e = 0)} - 1 \right) \quad (9)$$

which is equivalent to

$$TIL(U,\delta_e) = \left(\frac{T_{rotor}(U,\delta_e)}{T_{rotor,STrim}(U)} - 1\right) \times 100\% \quad (10)$$

We consider TIL as a function of flight speed and elevator deflection angle to measure the increment of rotor airloads with a certain $\delta_e$ at the velocity U, compared with the case without the elevator control. $T_{rotor}$ represents the main rotor airloads (and rotor thrust) under different strategy. $T_{rotor,STrim}(U)$ is equivalent to $T_{rotor}(U, \delta_e = 0)$. TIL serves as an indicator of the side effect results from using the elevator on the rotor. In this strategy, the trim problem is reformulated in (11).

$$\text{Given } U \in [0,100], \text{ Find } \delta_e = \arg\min_{\delta_e, \mathbf{x}} Power(\delta_e, \mathbf{x}) \text{ subject to } \begin{cases} \sum \mathbf{F(x)} = \mathbf{0} \\ \sum \mathbf{M(x)} = \mathbf{0} \\ \theta(U) = \theta_{preset}(U) \\ TIL(U,\delta_e) \leq \overline{TIL} \end{cases} \quad (11)$$

$\overline{TIL} \in [0,+\infty]$ is the upper bound of TIL. The exact value should be determined by the design of the material and the airfoil of the blade, and the rotational speed of the rotor. HTrim in this article takes a conservative value of 5% of $\overline{TIL}$. This is much less than 26.1% of TIL under MPTrim with an elevator deflection angle of -10° (see **Table 3**).

2.5.5 Remark

**Table 4** gives a summary of these four strategies. Note that the baseline BL does not have redundant trim problems, so the pitch angle preset strategy is only applicable for STrim, MPTrim, and HTrim, but not to the BL, whose pitch angle is directly derived from equation (5). Therefore, in the following comparison, it should be kept in mind that the pitch angles of STrim, MPTrim and HTrim are the same, but they are all different from the BL (especially at high speed). The optimization algorithms for MPTrim and HTrim will be explained in detail in the next section.

**Table 4**

Brief summary of different configurations

| Strategy | Propeller | Elevator |
|---|---|---|
| BL | Disabled | Disabled |
| STrim | Enabled | Disabled |
| MPTrim | Enabled | Enabled for minimum power |
| HTrim | Enabled | Enabled for minimum power with constraint of TIL |

*2.6 Optimization algorithm*

The required power, rotor airloads and TIL can be expressed as functions of the flight velocity U and the elevator deflection angle $\delta_e$. We define the corresponding function:

Objective function (required power):

$p:[0,100] \times [-15,0] \to \mathbb{R}$
$(U,\delta_e) \mapsto p(U,\delta_e)$

Main rotor load function:

$l:[0,100] \times [-15,0] \to \mathbb{R}$
$(U,\delta_e) \mapsto l(U,\delta_e)$

TIL function:

$TIL:[0,100] \times [-15,0] \to \mathbb{R}$
$(U,\delta_e) \mapsto TIL(U,\delta_e) = \left(\frac{l(U,\delta_e)}{l(U,\delta_e=0)} - 1\right) \times 100\%$

Several properties as preliminary information could be concluded from **Fig. 6**:

**Property 1**: For a given flight speed $U_0$, the function $l$ and $TIL$ decrease monotonically in $\delta_e \in [-15,0]$.

**Property 2**: For a given $U_0$, supposed there exists $\delta_e^* \in [-15,0]$ such that $\delta_e^* = \arg\min_{\delta_e} p(U_0,\delta_e)$, $p$ decreases monotonically in interval $\delta_e \in [-15, \delta_e^*]$, and increases monotonically in $\delta_e \in [\delta_e^*, 0]$.

Considering that [-15,0] is bounded, according to **Property 1** and **Property 2**, there exists one and only one local minimum of $p$, which is equal to the global minimum. Since the function $p$, $l$ and $TIL$ are nonlinear and not analytical, they could be calculated by numerical method.

For this nonlinear constraint optimization, a set of algorithms has been studied for many years such as gradient descent, sequential quadratic programming, L-M algorithm (the same as the one used for the trim). However, with the help of the properties, a more intuitionistic and high efficient algorithm called the heuristic descent search is used in this article. Since we know the variation of $p$, $l$ and $TIL$, we iteratively decrease the $\delta_e$ from 0 to -15 to approach the optimal power if the constraint is satisfied. Once the constraint is violated or the power starts to increase, the iterative procedure stops. The previous value of power in the iterations is recorded as the optimal power, since it is the last point which satisfies the $\overline{TIL}$ limitation with decreasing power. Moreover, it is both a local minimum and the global minimum (explained in the previous paragraph).

The main concern is the iterative descent rate of $\delta_e$. In order to be computationally efficient, we update the smallest search window each time when the program detects the minimum point. Then the descent rate will be divided by 10 to more precisely search the exact point of the minima. Besides, the length of the search window is automatically divided by 10. The algorithm begins with a descent rate of 1(deg) and stops until 0.01(deg). The advantage of this adaptive search window is to save the calculation time, because the update of $\delta_e$ is quick at the beginning and gradually slows down as p approaches the optima. The pseudocode in **Algorithm 1** provides more information about the calculation process.

**Algorithm 1** Heuristic descent search

**Input:** The CCH model, flight velocity $U_0$, elevator deflection angle range (deg) [-15,0], required power function $p$, rotor load function $l$, TIL function $TIL$.
**Output:** Optimal power $p^*$, optimal elevator deflection angle $\delta_e^*$.
1: $L \leftarrow -15$ (left bound of the search window)
2: $R \leftarrow 0$ (right bound of the search window)
3: $\alpha \leftarrow 1$ (step size)
4: $p^* \leftarrow +\infty$ (initialization)
5: **while** $\alpha \geq 0.01$ **do**
6:     $\delta_e = R$
7:     **while** $p(U_0,\delta_e) \leq p^*$ AND $TIL(U_0,\delta_e) \leq \overline{TIL}$ **do**
8:         $\delta_e \leftarrow \delta_e - \alpha$
9:         Trim the CCH model through L-M algorithm
10:         $p(U_0,\delta_e) \leftarrow$ required power
11:         $l(U_0,\delta_e) \leftarrow$ rotor load
12:         $TIL(U_0,\delta_e) \leftarrow \left(\frac{l(U_0,\delta_e)-l(U_0,0)}{l(U_0,0)} - 1\right) \times 100\%$
13:     **end while**
14:     $L \leftarrow \delta_e$
15:     $R \leftarrow \delta_e + \alpha$
16:     $\alpha \leftarrow \alpha/10$
17: **end while**
18: **return** $\delta_e^* = \delta_e + \alpha$, $p^* = p(\delta_e^*)$

**Algorithm 1** Heuristic descent search

This algorithm is applicable to both for MPTrim and HTrim, with $\overline{TIL} = +\infty$ for MPTrim (equivalent to non-TIL-constraint problem) and $\overline{TIL} = 5\%$ for HTrim.

**3. Results**

*3.1 Validation of the model*

The typical trim solution of STrim strategy is presented in **Fig. 8** from hover state to 100 m/s, showing a good agreement with those in the literature [13]. The CCH under STrim strategy and in the literature are both equipped similarly with coaxial rotors and pusher propeller, disabling the elevator control. The comparison results suggest that the flight dynamics model in this article is valid.

It also can be seen that the collective pitch in **Fig. 8**(a) is about 5 degrees higher than the reference trim at any flight speed, which may result from different setting angles of the rotor blades. In addition, the pusher propeller thrust begins to increase at 20 m/s in **Fig. 8**(f) and produces a huge thrust of about 8kN at 100 m/s, which makes a significant power consumed at high speed. A more detailed discussion of the trim solution under different strategies is presented in section 3.3.

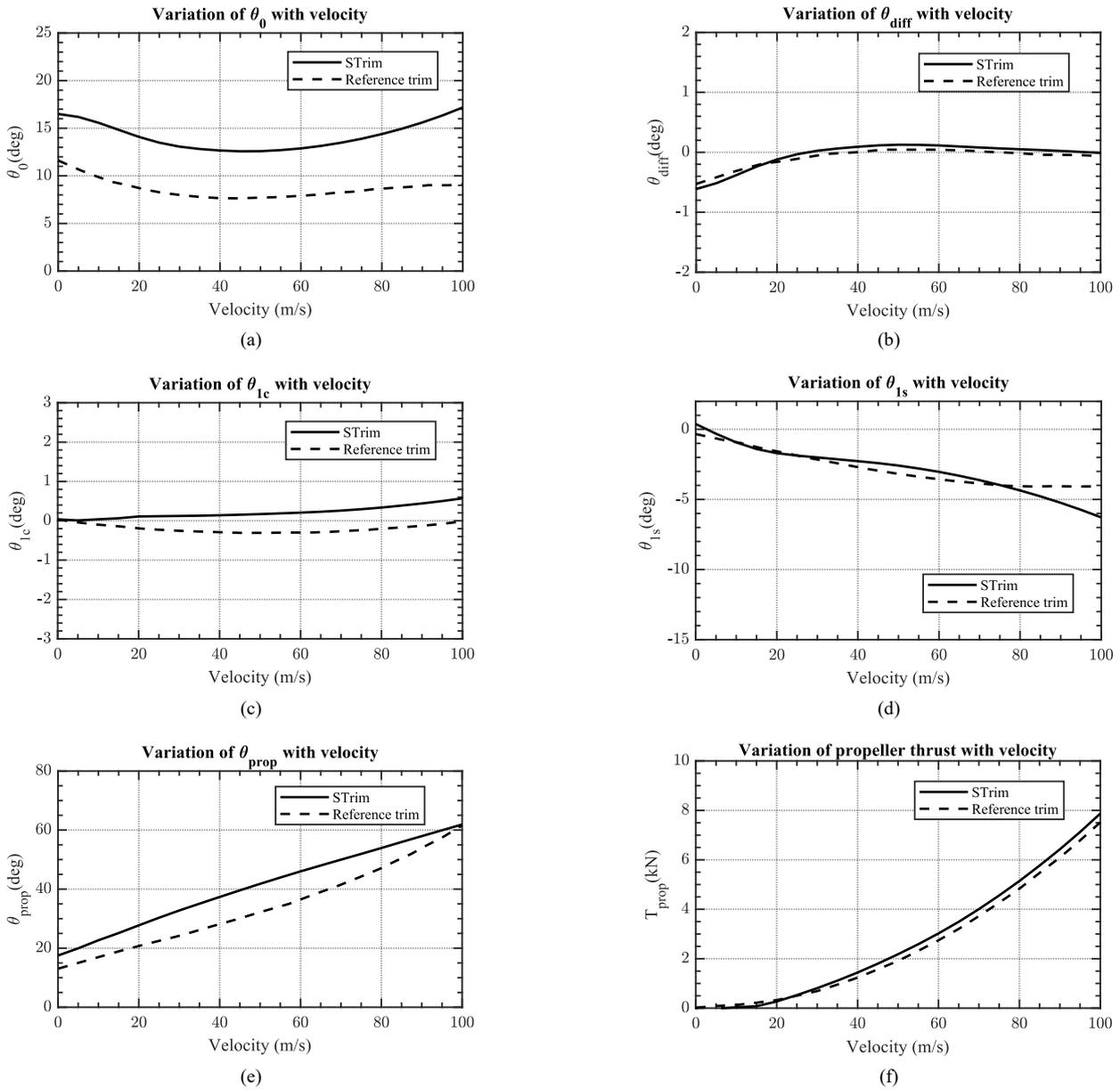

**Fig. 8** Validation of trim result

### 3.2 The optimal velocity to engage the propeller and the elevator

Since the pitch angle is pre-defined in **Fig. 4**, STrim, MPTrim, and HTrim all need to adjust the propeller thrust to maintain the pitch attitude, especially at high speed when the pitch angle is closer to zero compared to the conventional helicopter, while there is no need to engage the propeller at low speed.

The elevator control is evolved in MPTrim and HTrim with the equation (8) and the equation (11) providing the mathematical expressions for the optimal elevator deflection angle. It is foreseeable that the start of the elevator needs a flight speed at least 40-45m/s with higher aerodynamic efficiency (see **Fig. 5**). Moreover, it is hardly necessary to engage the elevator if the control efficiency is around zero at the flight speed below 40m/s.

**Fig. 9** demonstrates the variation of the required power with different speeds in 0-100m/s under the four trim strategies. The interval of calculation is 5m/s. Note that the range 70-100m/s for BL is only a theoretical extend since a conventional helicopter could not achieve such speed.

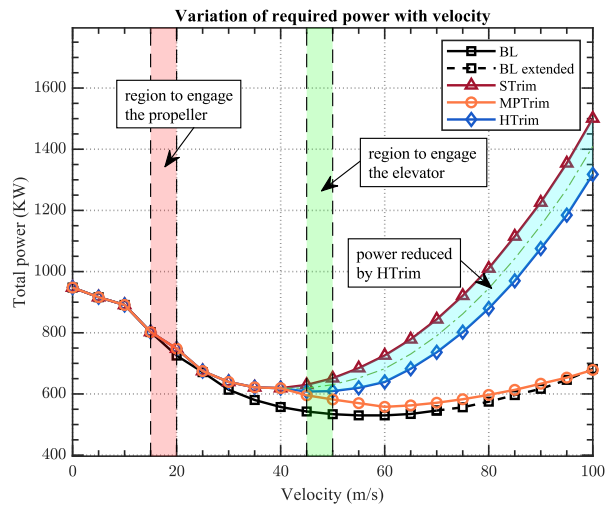

**Fig. 9** Required power with different velocities

The optimal speeds, or the optimal timings for starting the propeller and engaging the elevator control, represented as the red and green bands in **Fig. 9**, are respectively 15-20m/s and 45-50m/s, which also correspond to the separating points of the curves. **Table 5** is the resume of these two optimal velocities. Note that the red and green areas are not the speed range to 'use' the propeller and the elevator, but to 'start engaging them'. In other words, they are the minimum speeds to engage the propeller or the elevator, which also correspond to the turning points in **Fig. 11** and **Fig. 13**.

Table 5

Minimum velocities to engage the propeller or the elevator

| Minimum velocity (interval) to engage the propeller | Minimum velocity (interval) to engage the elevator |
|---|---|
| 15-20m/s | 45-50m/s |

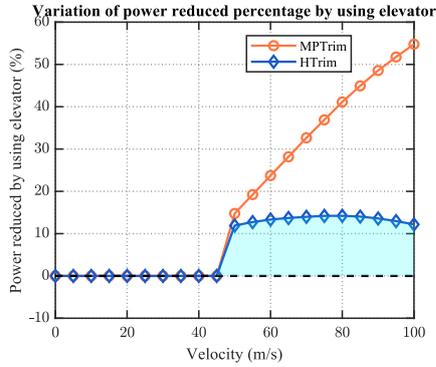

**Fig. 10** Power reduced percentage with different speed

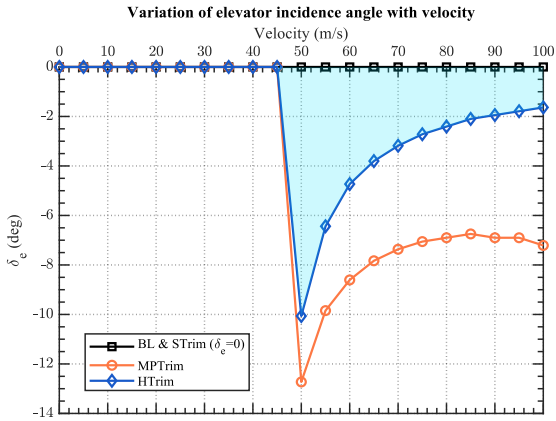

**Fig. 11** Elevator incidence angle with different velocity

Note that the required power under different strategies satisfies the following relationship:

$$\text{STrim} \geqslant \text{HTrim} \geqslant \text{MPTrim} \geqslant \text{BL} \qquad (12)$$

When flying below 15m/s, the required power of these four strategies is basically the same. According to **Fig. 10**, starting from 20m/s, the propeller begins to work (except for BL) and the power curves start to diverge. The introducing of the propeller requires extra power compared to BL, which becomes highly demanding as the speed increases.

The second separation of the curves begins at 45m/s, since the elevator of MPTrim and HTrim starts to be engaged according to **Fig. 11**. MPTrim searches the minimum possible power, while HTrim takes into account the main rotor airloads limitation, the $\overline{TIL}$. Since the required power is negatively correlated with the main rotor airloads, it is reasonable that HTrim could not achieve a more "economical" power than MPTrim but only reduces the power to a restricted extent. The cyan areas in **Fig. 9** and **Fig. 10** are the power reduction of HTrim compared to STrim. It can be found that the power is reduced by up to 14% with HTrim, and by 54% with MPTrim. The cyan area in **Fig. 11** represents the feasible range of elevator deflection angle satisfying the $\overline{TIL}$ constraint, whose corresponding power may not be necessarily the minimum (except on its boundary: the HTrim curve). Note that since the aerodynamic of the elevator significantly increases with the flight speed, a higher speed requires a smaller deflection angle.

In addition, we noticed that MPTrim approaches the BL curve at 100m/s. This is because the optimal solution found by MPTrim corresponds a zero thrust of the propeller, which is equivalent to disabling the propeller. This could be verified from **Fig. 13**, where the propeller thrust above 50m/s is near 0 for MPTrim.

**Fig. 12** illustrates the increase of rotor airloads as a side effect of using the elevator. We should notice that, although MPTrim finds the minimum power, it also brings a tremendous increase in rotor airloads, which may result in issues such as vibration. As an optimal control strategy, HTrim does not consider the minimum power as the only indicator but searches for a compromise between the constraints of power and main rotor airloads. In this way, while reducing the power to a certain extent, the airloads of the main rotor would always satisfy the limitation of TIL.

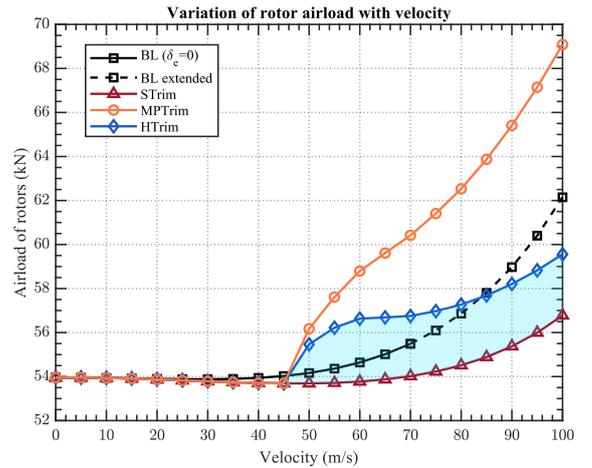

**Fig. 12** Variation of rotor airload with different velocity

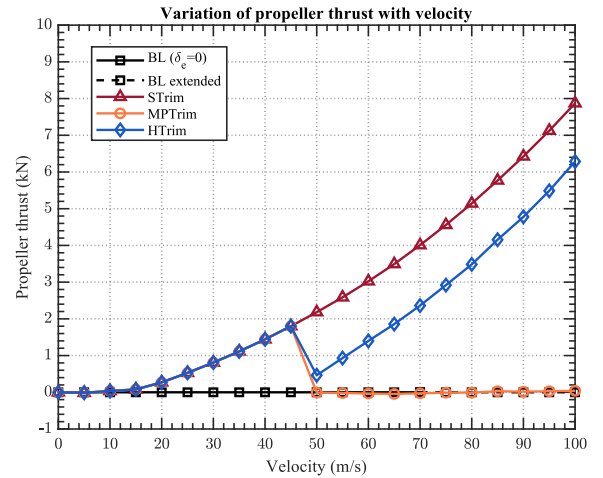

**Fig. 13** Variation of propeller thrust with different velocity

This paper uses 5% as a conservative value for the maximum TIL: $\overline{TIL}$. For other rotorcraft with different configurations, $\overline{TIL}$ should be considered comprehensively according to the design parameters.

### 3.3 Results of the main controls and the flight attitude

This section compares the impact of different elevator strategies on the attitude and the main control variables in Figure 13.

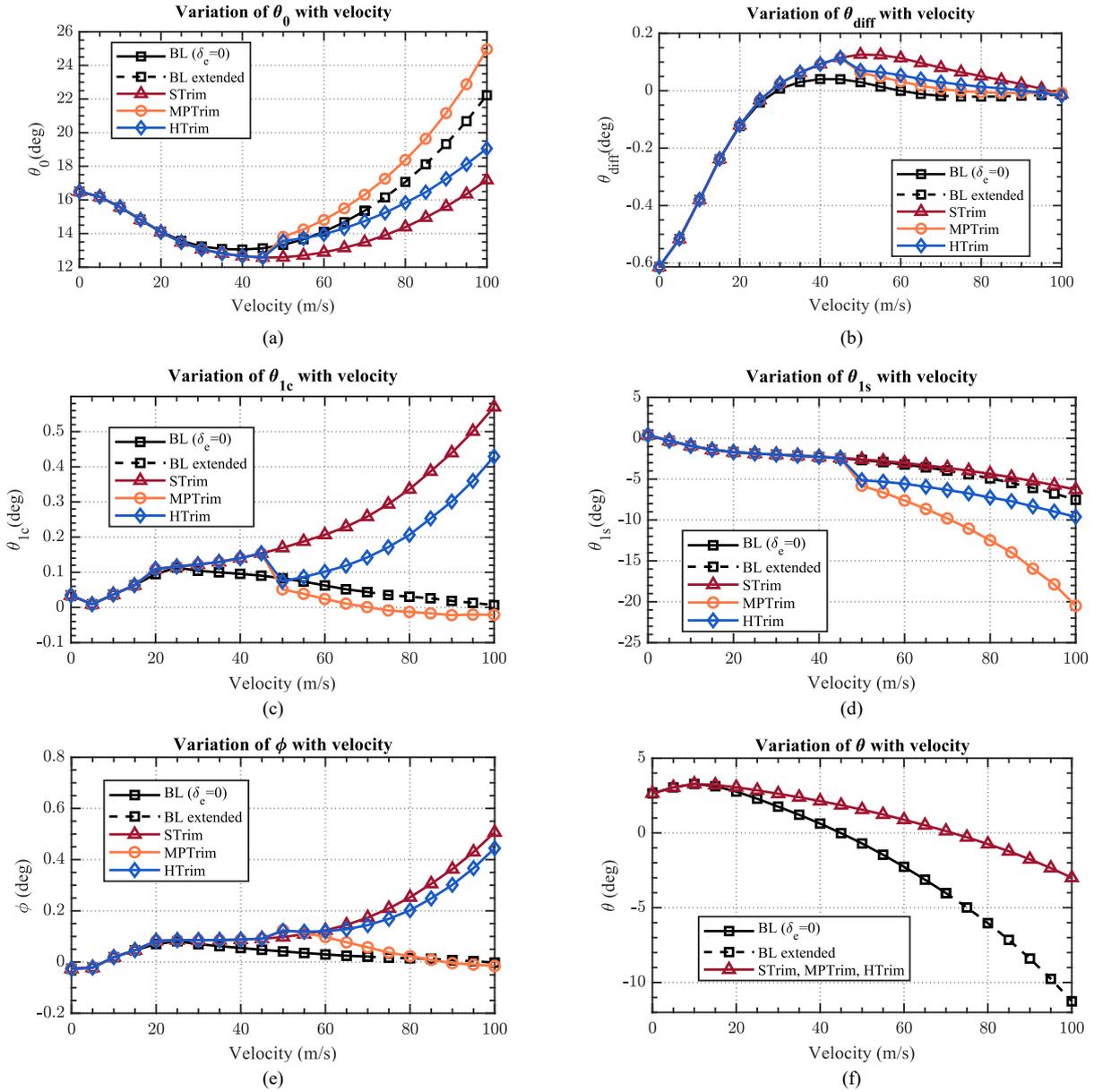

Fig. 14 Trimmed results of main control variables

Each curve corresponds to the results obtained by using different elevator control strategies. The following points are worth noting.

1. Since the pusher propeller and the elevator controls are engaged at 15-20m/s and 45-50m/s, the curves begin to separate around these speeds.

2. Note that compared to BL, STrim requires the smallest collective pitch, implying that the main rotor airloads is the smallest. The elevator is disabled by STrim, and the pusher propeller is maximally used to balance the drag generated by the fuselage, so the main rotor airloads can be unloaded to the greatest extent. On the contrary, MPTrim almost shuts down the propeller above 50m/s and the main rotor needs to generate more forward force to replace it with an increased rotor airloads. HTrim is a compromise between STrim and MPTrim.

3. The tendencies of the differential collective are basically the same with convergence to 0 at 100m/s. The differential collective is mainly used to balance the yaw moments produced by the upper and lower rotors. At high speeds, the wakes of the upper and lower rotors are almost parallel to the rotor disk plane, leading to a more similar aerodynamic characteristic of rotors. Therefore, the generated forces and moments are likely to have the same value, so little manipulation of differential collective is required to balance them.

4. The lateral cyclic control is for balancing the roll moments of the upper and lower rotors at low speed, and for balancing the propeller torque at high speeds. Since the STrim produces the maximum propeller thrust and torque among these strategies, the magnitude of lateral cyclic pitch is also the largest. In addition, note that the lateral cyclic control is related to the roll angle of the aircraft, the curves of which are also highly similar.

5. The longitudinal cyclic control is the main manipulation to generate pitch moments and forward thrust of main rotors. It can be seen in **Fig. 14** (d) that the BL without propeller (and MPTrim above 50m/s) needs large longitudinal cyclic pitch at high speed to provide forward thrust to balance the enormous drag generated by the fuselage. Pusher propeller can significantly reduce this manipulation represented as the STrim (compared to BL) and HTrim (compared to MPTrim) in **Fig. 14** (d). It should be noted that although BL and STrim almost overlap, their corresponding pitch angles (**Fig. 14** (f)) are different. For STrim, it is unacceptable that a pilot would find a lower pitch angle at higher velocities. This problem does not exist under other control strategies.

6. The adaptability of the optimal strategy HTrim could be identified through the optimal speed range to engage the propeller and elevator at respectively 15-20m/s and 45-50m/s. The control varies between the STrim and MPTrim in a desired controllable range with a relatively smooth variation.

7. Combining **Fig. 9**, **Fig. 11**, and **Fig. 13**, the MPTrim above 50m/s behaves very similar to BL which disengages the propeller, and the forward thrust is completely provided by the main rotors. The difference between MPTrim and BL is that the zero-thrust of propeller under BL is preset but not for MPTrim, which searches for a possible mathematical reduction of the required power to a maximum extent. Since the propeller consumes a significant power at high speed, MPTrim tries to avoid using it without considering other constraints such as the main rotor airloads. Furthermore, as BL is not achievable at high speed, neither could MPTrim be realistic for high-speed control for the same reasons. However, it could still be a reference for low-speed or medium-speed flight.

Once again, the manipulations of STrim and MPTrim are the two extremes of the elevator control strategy (see **Fig. 11**). Moreover, the MPTrim is not achievable at high speed because of a severe increase in blade airloads and the catastrophic consequences. HTrim is a compromise of the two, that is, the reduction of power is realized accompanied by an increase of rotor airloads limited within 5%. In this case, HTrim could be regarded as an optimal strategy for elevator control.

### 4. Discussion

Compared to the existing work[22], where the author investigates the control derivatives, bandwidth and phase delay corresponding to longitudinal cyclic control and elevator, we further study the effect of elevator control on trim operation, force distribution across components, required power, and rotor airloads. On this basis, we propose an optimized longitudinal control strategy based on required power and rotor airloads to improve the flight performance of CCH at medium and high speed. The control allocation of the elevator and longitudinal pitch is derived from a nonlinear optimization based on the performance indices (power and rotor airloads), while the strategy in the literature[22] follows a fixed heuristic equation, where the allocation of the elevator increases quadratically with airspeed.

Although we define the performance function in HTrim strategy, it is more difficult to clearly define the allocation relation of HTrim (between the elevator and the longitudinal cyclic pitch) by an explicit analytical formula, compared to the strategy in the literature[22], where the implementation of the control strategy is more feasible from an engineering perspective. This is because the result of HTrim is obtained by a complex nonlinear numerical optimization, rather than adding an explicit constraint to the trim equation system. However, the HTrim strategy still has huge application potential, especially in the scenarios based on performance function requirements with nonlinear constraints (such as the maximum rotor airloads).

Therefore, the practical application of HTrim still faces great challenges. But we believe that the continuous development of computer technology and flight control theory could eventually overcome this problem, which is also an important part of our further research work.

### 5. Conclusion

The article starts with the establishment of a validated CCH model, focusing on the analysis of the effect from the elevator and the propeller on trim, then compares different trim strategy and proposes HTrim that optimizes the performance in required power and rotor airloads. The main conclusions are as follows:

1. The elevator has little aerodynamic efficiency below 40m/s, which gradually increases from 45m/s and above. At medium and high speeds, the influence of the elevator on trim control is significant.
2. The primary influence of the elevator is on the total airloads of the rotor, the distribution of roll moment, the load of the pusher propeller, and the required power of the CCH. The deflection angle of the elevator is positively correlated to the required power and the propeller thrust, while negatively correlated to the rotor airloads and the roll moments of the upper and lower rotor. If the rotor airloads limitation is not taken into consideration, the choice of elevator control strategy can reduce the required power by as much as 54.2%, with an increase of rotor airloads by 26.1% (at 100m/s). This strategy is not realistic for real flight (especially at high speed) because of the severe problems brought by the increase of the rotor airloads: such as the retreating blade stall, advancing blade compressibility and the vibration problem.
3. Considering the limit of rotor load, the concept of Toleration of Increased Load (TIL) is proposed. By setting the upper bound of TIL at 5%, the Hybrid Trim strategy is established, under which the elevator can reduce the required power by up to 15% with an increase of rotor load limited below 5%. Compared with Simple Trim (i.e., without the elevator control), it will slightly increase the main rotor collective control and reduce the collective differential, longitudinal/lateral cyclic pitch, and propeller thrust. With a careful selection of the upper bound value (5% in this article, but could be other value in real design) of TIL according to the CCH parameters and design, the Hybrid Trim strategy can be recommended as an optimal strategy for real flight.
4. The engaging order of the propeller and the elevator under the Hybrid Trim strategy is established, with the optimal (also the minimum) speed to enable the propeller and elevator at respectively 15-20m/s and 45-50m/s.

It is worth commenting that, the choice of the control strategy could result in loss or gain of dynamic range of controllability and maneuverability. In addition, the optimal solution also faces the challenge of implementation in the flight control system since the control allocation results is implicit, which is the main focus of our follow-up work.